\documentstyle[epsfig,12pt]{article}
\newcommand{\la}[1]{\label{#1}}
\newcommand{\ur}[1]{(\ref{#1})}

\newcommand{\eq}[1]{eq.~(\ref{#1})}
\newcommand{\eqs}[2]{eqs.~(\ref{#1},\ref{#2})}
\newcommand{\Eq}[1]{Eq.~(\ref{#1})}
\newcommand{\Eqs}[2]{Eqs.(\ref{#1}, \ref{#2})}

\newcommand{\e}{\epsilon}

\newcommand{\half}{\frac{1}{2}}

\def\Tr{\mbox{Tr}}
\def\det{\mbox{det}}

\def\beq{\begin{equation}}
\def\eeq{\end{equation}}
\def\bea{\begin{eqnarray}}
\def\eea{\end{eqnarray}}

\begin{document}
\thispagestyle{empty}
\begin{flushright} NORDITA-2002-79-HE
\end{flushright}
%\begin{flushright} 16 Dec 2002
%\end{flushright}
%\begin{flushright} .....
%\end{flushright}
%\today
\vskip 2true cm
\begin{center}
{\Large\bf Yang--Mills theory in terms of gauge invariant \\
\vskip .4true cm

dual variables}

\vskip 1.5true cm

{\large\bf Dmitri Diakonov$^{\diamond *}$} 
\vskip .8true cm
$^\diamond$ {\it NORDITA, Blegdamsvej 17, DK-2100 Copenhagen \O,
Denmark} \\
\vskip .5true cm
$^*$ {\it St. Petersburg Nuclear Physics Institute, Gatchina 188 300,
Russia} \\
\vskip .2true cm
%E-mail: diakonov@nordita.dk
\end{center}
\vskip 1true cm
\begin{abstract}
\noindent
Quantum Yang--Mills theory and the Wilson loop can be rewritten identically in
terms of local gauge-invariant variables being directly related to the metric of the
dual space. In this formulation, one reveals a hidden high local symmetry of the
Yang--Mills theory, which mixes up fields with spins up to $J=N$ for the $SU(N)$
gauge group. In the simplest case of the $SU(2)$ group the dual space seems to 
tend to the de Sitter space in the infrared region. This observation suggests a new
mechanism of gauge-invariant mass generation in the Yang--Mills 
theory \footnote{Invited talk at {\it Confinement-5}, Lago di Garda, 10-14 Sep 2002.}.
\end{abstract}

\vspace{.7cm}
\section{Introduction}

Ever since the formulation of Quantum Chromodynamics
and the realization that only gauge-invariant operators are the observables,
there have been attempts to reformulate the theory itself not in terms of
the Yang--Mills potentials but rather in terms of some gauge-invariant variables.
However, usually it is difficult to incorporate matter in such approaches
as it couples not to gauge-invariant variables but to the gauge potentials
$A_\mu$.

The simplest but typical matter is the probe source, or the Wilson loop.
It is given by a path-ordered exponent of $A_\mu$, and it is difficult
to present it in terms of any gauge-invariant variables. At some point
it has been suggested that Wilson loops themselves are the only
reasonable gauge-invariant variables, and the theory should be
reformulated in terms of loop dynamics \cite{AMP1,MM}. Because Wilson
loops are non-local objects there have been no decisive success on this
way, despite enormous efforts taken in twenty years.

Recently, it became clear that it is not only possible to reformulate exactly the
YM theory (together with the Wilson loop) in terms of {\em local} gauge-invariant
variables, but that the resulting theory is beautiful and suggestive. It reminds
quantum gravity theory, where the metric and curvature refer to the {\em dual}
YM space (below I specify what precisely does it mean). It becomes possible
to discuss the long-standing but still unresolved questions like the mass generation and the
area behaviour of the Wilson loop directly in a gauge-invariant, {\it i.e.} objective way. Also,
the common belief is that, at least at large number of colours $N$, the YM theory is
equivalent to some version of string theory. String theory is formulated in terms of
gauge-invariant although non-local variables. Recent attempts to justify this equivalence from
the string side have been described by A. Polyakov \cite{AMP2}. If one wishes to derive
string theory directly from the YM Lagrangian, the first step is to rewrite the theory in terms of
gauge-invariant variables. I shall show below that the YM theory possesses an exciting new
symmetry mixing states with different spins,  which makes the equivalence with string theory
rather likely.

\section{First order formalism}
\subsection{Four dimensions}

An economic way to introduce gauge-invariant variables is via the so-called
first order formalism. The $4d$ YM partition function can be identically
rewritten with the help of an additional Gaussian integration over dual
field strength variables \cite{DT,Halpern}:
\bea
\nonumber
Z_{4{\rm d}} &=& \int DA_\mu\,\exp\int d^4x\;\left(- \frac{1}{2g_4^2}\Tr\,
F_{\mu\nu}F_{\mu\nu}\right)\\
&=&\!\!\! \int\!\! DA_\mu\,DG_{\mu\nu}\,\exp\int\!
d^4x \left(\! -\frac{g_4^2}{2}\Tr\, G_{\mu\nu}G_{\mu\nu}
+\frac{i}{2}\,\epsilon^{\alpha\beta\mu\nu}\,\Tr\,G_{\alpha\beta}F_{\mu\nu}\!\right)
\la{Z41} \eea
where $F_{\mu\nu}=\partial_\mu A_\nu-\partial_\nu A_\mu-i[A_\mu A_\nu]$
is the standard Yang--Mills field strength and
$\epsilon^{\alpha\beta\mu\nu}$
is the antisymmetric tensor.  To be specific, the gauge group is $SU(N)$ with
$N^2\!-\!1$ generators~$t^a$, $\Tr\,t^at^b=\delta^{ab}/2$.  \Eq{Z41} is
called the 1st order formalism.

Both terms in \eq{Z41} are invariant under $(N^2\!-\!1)$-function gauge
transformation,
\beq
\left\{\begin{array}{ccc} \delta A_\mu &=&
[D_\mu\, \alpha],\\ \delta G_{\mu\nu} &=& [G_{\mu\nu}\,\alpha]
\end{array}\right.
\la{GT4}\eeq
where $D_\mu=\partial_\mu -iA^a_\mu t^a$ is the Yang--Mills
covariant derivative, $[D_\mu D_\nu]=-iF_{\mu\nu}$.

Owing to the Bianchi identity,
$\epsilon^{\mu\nu\rho\sigma}\,[D_\nu F_{\rho\sigma}]=0$,
the second (mixed) term in \eq{Z41} is, in addition, invariant under the
$4(N^2-1)$-function `dual' gauge transformation,
\beq
\left\{\begin{array}{ccc} \delta A_\mu &=& 0,\\
\delta G_{\mu\nu} &=&[D_\mu\beta_\nu]-[D_\nu\beta_\mu].
\end{array}\right.
\la{DGT4}\eeq
Taking a particular combination of the functions in
\eqs{GT4}{DGT4},
\beq
\alpha=v^\mu A_\mu,\qquad \beta_\mu=v^\lambda
G_{\lambda\mu}, 
\la{CT}\eeq 
leads to the 4-function transformation 
\beq 
\delta G_{\mu\nu}= -G_{\lambda\nu}\,\partial_\mu v^\lambda
-G_{\mu\lambda}\,\partial_\nu v^\lambda -\partial_\lambda
G_{\mu\nu}\,v^\lambda,
\la{CTT}\eeq
being the known variation of a (covariant) tensor under infinitesimal general
coordinate transformation, $x^\mu\to x^\mu+v^\mu(x)$, also called the
diffeomorphism. Therefore, the `mixed' term is
diffeomorphism-invariant, and is known as BF gravity~\footnote{With our
notations it would be more appropriate to call it `GF gravity' but we follow
the tradition.}. It defines a topological field theory of the Schwarz type
\cite{BFobzor}. Moreover, it is invariant not under four but as much as
$4(N^2\!-\!1)$ local transformations; four diffeomorphisms are but their
small subset. We shall see later on that the additional local transformations
mix up fields with different spins.

In the first order formalism \ur{Z41}, the YM potential $A_\mu$ enters
linearly and quadratically. Therefore, on can now perform the Gaussian
integration over $A_\mu$: the quadratic form for $A_\mu$ is, generally,
non-degenerate and contains no derivatives. Thus, one has just a product
of Gaussian integrals at each point, and it is not too difficult to
perform the $A_\mu$ integration explicitly~\cite{DT,Halpern}. The point is
to write the result in a nice way such that, for example, the
diffeo-invariance is explicit. This has been done in Ref.~\cite{GS},
however the result there is more than one page long, and the invariance
under $4(N^2\!-\!1)$-function transformation \ur{DGT4} has been neither
revealed nor discussed. It has been achieved in a compact way in 
Ref.~\cite{DPBF} which we review in the next section.

\subsection{Three dimensions}

We shall be also interested in the $3d$ YM theory: it is also believed to
possess the area law for large Wilson loops and an exponential decrease
of correlators of small Wilson loops at large separations. The theory
is super-renormalizable; the mass gaps and string tension are proportional
to the gauge coupling $g_3^2$ (having the dimension of mass) in the
appropriate power.

In the $3d$ case one can also use the Gaussian trick to rewrite the partition
function in the first order formalism. In this case the dual field strength
is a 3-vector, call it $G^a_i$ (the Latin indices go from 1 to 3).
\bea
\nonumber
Z_{3{\rm d}}&=& \int DA_i^a\, \exp\left( -\frac{1}{4g_3^2}\int d^3x\,
F_{ij}^a(A)F_{ij}^a(A)\right)\\
\la{Z31}
&=&\int\!
DG_i^a\,DA_i^a\,\exp\int\!d^3x\left[-\frac{g^2_3}{2}G_i^aG_i^a
+\frac{i}{2}\e^{ijk}F_{ij}^a(A)G_k^a\right].
\eea

As in the $4d$ case, the Bianchi identity, $\e^{ijk}D^{ab}_iF^b_{jk}=0$,
ensures that the second (mixed) term in \eq{Z31} is invariant under
local $(N^2-1)$-function dual gauge transformation
\beq
\left\{\begin{array}{ccc}
\delta A^a_i&=&0,\\
\delta G^a_i&=&D^{ab}_i(A)\,\beta^b\end{array}\right..
\la{DGT3}\eeq
Again, the invariance under ordinary gauge transformation together with the dual
one guarantees that, if one integrates out the YM potentials $A_i$, the
resulting action will be invariant under local $(\!N^2-1\!)$-function
transformations, three of which are the diffeomorphisms. For $SU(2)$
that is the only invariance but at $N>2$ the invariance is much wider
than that of the general relativity.

\section{Yang--Mills theory as quantum gravity with `\ae ther'}

In this section we give the results of integrating out the YM potential
$A_\mu$ from the first-order formalism's \eqs{Z41}{Z31}. The goal is
to obtain the result in terms of gauge-invariant combinations
of the dual field strength $G_{\mu\nu},\,G_i$ and to see that
the second (mixed) term has the claimed large local invariance.

\subsection{Three dimensions, $SU(2)$}

We start with the simplest case of the $SU(2)$ gauge group in $3d$.
In this case one can identify the dual field strength $G^a_i=e^a_i$
with a covariant {\it dreibein}~\cite{Lun1}. Integrating out $A^a_i$
in \eq{Z31} one obtains, identically \cite{GS,AMS,DP1,DP2}

\beq
Z_{3{\rm d}}=\int Dg_{ij}\,g^{-\frac{5}{4}}
\exp\int\!d^3x\,\left[-\frac{g_3^2}{2}\, g_{ii}+\frac{i}{2}\sqrt{g}R(g)
\right].
\la{Z32}\eeq
where $g_{ij}=e^a_ie^a_j$ is the covariant metric tensor of the dual
space constructed, as we see, from the dual field strength. Knowing the
metric $g_{ij}$ one can build the Christoffel symbol $\Gamma^i_{jk}$,
the Riemann tensor $R^i_{jkl}$, the Ricci tensor $R_{ij}$ and finally
the scalar curvature $R$ according to the general formulae of differential
geometry, see {\it e.g.} \cite{DP3}; $g$ is the determinant of $g_{ij}$.

The second term in the action \ur{Z32} is the familiar Einstein--Hilbert
action in $3d$; notice that it comes with a purely imaginary `Newton
constant'. The fact that the second term is invariant under the
diffeomorphisms of the dual space is the consequence of the invariance
of the mixed term in the first order formalism under dual gauge
transformations. The first term in the action, $g_{ii}$, is just a
rewriting of the first term from \eq{Z31}; it is not invariant under
dual gauge transformations and hence it is not invariant under
diffeomorphisms. We call it the `\ae ther term' as it reminds the coupling
of gravity to matter with the stress-energy tensor $T^{ij}=\delta^{ij}$
which is isotropic and homogeneous but makes one system of coordinates
preferable. It is this \ae ther term which distinguishes the YM theory
from pure Einstein gravity. In $3d$ Einstein's gravity is a 
non-propagating topological field theory \cite{Wit}. \\

There are many alternative ways how to parametrize a curved space;
defining the metric $g_{ij}$ is only one of them, and not the best in this
case. One can define a curved $d$-dimensional space by (locally)
embedding it into a flat $D=d(d+1)/2$-dimensional space. In this case, the
flat space is $6d$. 6 functions $w^\alpha(x^1,x^2,x^3),\,\alpha=1...6,$
called the {\em external coordinates} define the embedding of a generic
$3d$ curved manifold into a flat $6d$ space. The metric is induced
by this embedding:

\beq
g_{ij}=\partial_iw^\alpha\partial_jw^\alpha.
\la{gw}\eeq
All differential geometry's quantities including the scalar curvature $R$
can be expressed through the six functions $w^\alpha(x)$ \cite{DP1}.
Invariance under diffeomorphisms is invariance under the
re-parametrization: $w^\alpha(x)\to w^{\prime\,\alpha}(x)$. In terms
of the external coordinates the YM partition function takes the form
\cite{DP1,DP2}
\bea
\nonumber
Z_{3{\rm d}}&=&\int Dw^\alpha(x)\,\det(\partial_i\partial_jw^\alpha)\,
g(w)^{-\frac{5}{4}}\\
&\cdot &
\exp\int\!d^3x\,\left[-\frac{g_3^2}{2}\,\partial_iw^\alpha\partial_iw^\alpha
+\frac{i}{2}\sqrt{g(w)}\;R(w)\right],
\la{Z33}\eea
so that the \ae ther term becomes just the kinetic energy of 6 massless
and gauge-invariant fields; however the curvature term brings in the
interaction.

In the original formulation of the $SU(2)$ YM theory in $3d$ there
were 9 degrees of freedom (dof's) in the gauge potentials $A^a_i$ out
of which 3 were gauge transformations. The remarkable achievement
of \eqs{Z32}{Z33} is that the partition function is presented in terms
of exactly 6 gauge-invariant variables: $g_{ij}$ or $w^\alpha$.
Averages of Wilson loops, since they are gauge-invariant, can be also
presented in terms of the gauge-invariant parametrization of the dual
space. It turns out \cite{DP2,DP3} that Wilson loops become, in the
`gravity' formulation, {\it parallel transporters} in the curved dual
space!

\subsection{Three dimensions, general $SU(N)$}

In the $3d$ $SU(N)$ YM theory there are $3(N^2\!-\!1)$ dof's in the
original formulation via the gauge potentials $A_i^a$, out of which
$N^2\!-\!1$ are gauge transformations, therefore, the number of
gauge-invariant dof's is $2(N^2\!-\!1)$.

As in the $SU(2)$ case, we wish to integrate out the YM potentials
$A^a_i$ and express the result in terms of gauge-invariant combinations
of the dual field strength $G^a_i$. In the $SU(2)$ case the $2\cdot 3=6$
gauge-invariant variables are quadratic combinations, $g_{ij}=G^a_iG^a_j$;
the metric tensor can be decomposed into spin 0 (1 dof) and spin 2 (5 dof's)
fields, $1+5=6$.

For $SU(3)$ one needs $2\cdot 8=16$ gauge-invariant dof's; these are the 6
components of the metric tensor $g_{ij}$, plus 10 components of spin 1
(3 dof's) and spin 3 (7 dof's) fields, put together into a symmetric
rank-3 tensor $h_{ijk}=\frac{1}{3}\Tr\{G_iG_jG_k\}$ where $G_i=G^a_it^a$ and
$\{\}$ means full symmetrization. In $SU(2)$ $h_{ijk}$ is automatically zero.
In $SU(4)$ one would need a rank-4 symmetric combination of $G_i$'s which
is automatically zero in $SU(3)$, and so on \cite{DPBF}.

The general pattern is that for the $SU(N)$ gauge group
one adds new spin $N$ and spin $N\!-\!2$ fields to the `previous' fields
of the $SU(N\!-\!1)$ group. All in all, one has for the $SU(N)$ two copies
of spin $2,3\ldots N\!-\!2$ and one copy of the `edge' spins 0, 1, $N\!-\!1$ and
$N$; they sum up into the needed $2(N^2\!-\!1)$ dof's \cite{DPBF}.
In the usual formulation, one gets $2(N^2\!-\!1)$ dof's just because
there is a colour index which runs up to $N^2\!-\!1$, however the presence of
a colour index implies the quantity is not gauge-invariant. If one wishes
to pass to gauge-invariant variables, it is inevitable that the `coloured'
dof's are `squeezed' into higher and higher spin fields.

The invariance of the second term in the first-order-formalism action
\ur{Z31} under $(N^2\!-\!1)$-function dual gauge transformation \ur{DGT3}
translates into $(N^2\!-\!1)$-function local symmetry which mixes fields with
different spins \cite{DPBF}. This exciting new symmetry can be paralleled
to that of supergravity where one can locally mix up spin 2 and spin $3/2$.
However, in our case only boson fields are mixed by symmetry and one can
mix up the whole tower of spins up to $J=N$. The transformation is non-linear,
though, so there is no contradiction with the Coleman--Mandula theorem.

After integrating out $A_i^a$, the second term reminds the Einstein--Hilbert
action but contains not only the metric tensor $g_{ij}$ but also higher
spin fields $h_{ijk}\ldots$. Since $SU(2)$ is a subgroup of any Lie group,
this part of the action is diffeomorphism-invariant but for $N>2$ it is
in addition invariant under other local transformations mixing fields
with different spins \cite{DPBF}. At $N\to\infty$ an infinite tower
of spins are related by symmetry transformation. Since it is the kind
of symmetry known in string theory, and some kind of string is expected to be
equivalent to the Yang--Mills theory in the large $N$ limit, it is
tempting to use this formalism as a starting point for deriving a
string from the local field YM theory. The \ae ther term $g_3^2N\,g_{ii}$
breaks this symmetry. Therefore, one expects that the role of this
term is to lift the degeneracy of the otherwise massless fields and to
provide the string with a finite slope $\alpha^\prime=(g_3^2N)^{-2}$.

\subsection{Four dimensions, $SU(2)$}

In $4d$ the dual field strength is an antisymmetric tensor,
not a vector, so to construct the metric tensor one has to be more
industrious. The needed metric tensor can be constructed as \cite{GS}

\beq
g_{\mu\nu}=\frac{1}{6}\epsilon^{abc}\,\frac{\epsilon^{\alpha\beta\rho
\sigma}} {2\sqrt{g}}\,G^a_{\mu\alpha}G^b_{\rho\sigma}G^c_{\beta\nu}.
\la{metr41}\eeq
However, it is not a convenient variable. A more suitable choice is
to present the dual field strength in \eq{Z41} as \cite{DPBF}

\beq
G^a_{\mu\nu}=d^a_i\,T^i_{\mu\nu}=d^a_i\,\eta^i_{AB}\,
e^A_\mu\,e^B_\nu,
\la{Gmunu}\eeq
where $\eta^i_{AB}$ is the 't Hooft symbol projecting the $(1,0)+(0,1)$
representation of the Lorentz $SO(4)$ group into the irreducible $(1,0)$ part.
$e^A_\mu$ can be called the tetrad, and the metric tensor is, as
usually, $g_{\mu\nu}=e^A_\mu e^A_\nu$. There are 16 dof's in the tetrad,
however three rotations under one of the $SO(3)$ subgroups of the $SO(4)$
Lorentz group does not enter into the combination \ur{Gmunu}, therefore,
the antisymmetric tensor $T^i_{\mu\nu}$ carries 13 dof's.

The $3\times 3$ tensor $d^a_i$ can be called the triad; it is subject
to the normalization constraint $\det\,d^a_i= 1$ and therefore contains
8 dof's. In fact, the combination \ur{Gmunu} is invariant under
simultaneous $SO(3)$ rotations of $T^i$ and $d_i$, therefore the r.h.s.
of \eq{Gmunu} contains $13+8-3=18$ dof's, as does the l.h.s. Thus,
\eq{Gmunu} is a complete parametrization of $G^a_{\mu\nu}$.

It is now clear how to organize the 15 gauge-invariant variables made of
$G^a_{\mu\nu}$. These are the 5 dof's contained in a symmetric $3\times
3$ tensor
\beq
h_{ij}= d^a_i\,d^a_j,\qquad \det\, h=1,
\la{defh}\eeq
and 13 dof's of $T^i_{\alpha\beta}$. However, $h_{ij}$ and
$T^i_{\alpha\beta}$ will always enter contracted in $i,j$ (as it
follows from \eq{Gmunu}), so that the dof's associated with the
simultaneous $SO(3)$ rotation will drop out. In other words, one can
choose $h_{ij}$ to be diagonal and containing only 2 dof's. Together
with the 13 dof's of $T^i_{\mu\nu}$ they comprise the needed 15
gauge-invariant degrees of freedom.

After integrating out $A_\mu$ from the first-order-formalism partition
function \ur{Z41} and expressing the result through $T$ and $h$ one
obtains the $4d$ YM partition function in terms of gauge-invariant
variables \cite{DPBF}:
\bea
\la{Z42}
Z_{4{\rm d}}&=&\int Dh DT \,e^{\,S_1+S_2},\\
\la{S1}
S_1&=&-\frac{g_4^2}{4}\int d^4x\, T^i_{\mu\nu}\,h_{ij}\,T^j_{\mu\nu},\\
\la{S2}
S_2&=&\frac{i}{4}\int d^4x\,\sqrt{g}\,R^j_{\;i\,\mu\nu}\,T^{l\,\mu\nu}\,
\epsilon_{jlk}\,h^{ki},
\eea
where $R^j_{\;i\,\mu\nu}$ is a `minor' Riemann tensor, see Ref. \cite{DPBF}
for details. We see that \eq{S2} is covariant both with respect to Greek
and Latin indices: it is the manifestation of the invariance of this
part of the action under a 12-function {\em local} transformation following
from the invariance under dual gauge transformations \ur{DGT4}. Four of these
local transformations are diffeomporphisms; the rest mix up spin 0 and
spin 2 fields. It is a `more general relativity' theory. However, in the particular case when
$h_{ij}=\delta_{ij}$ the action $S_2$ reduces to the usual Einstein--Hilbert action,
$\sqrt{g}R$, where $R$ is the standard scalar curvature made of $g_{\mu\nu}$. \\

I would like to note as an aside that one can think of making quantum gravity 
renormalizable in the following spirit. Start from the `more general relativity' theory 
given by the action $S_2$. It is not only renormalizable but probably exactly
solvable. If, for some reason, the field $h_{ij}$ develops spontaneously 
a v.e.v. $\sim \delta_{ij}$ one gets, at `low' energies the Einstein's quantum gravity, 
but renormalizable in the ultra-violet.

\section{Dual transformation on the lattice}

The duality transformation to gauge-invariant variables can be done
directly on the lattice -- see Refs.~\cite{Aetal,DP1} for $3d$ $SU(2)$, Ref. ~\cite{HS}
for $4d$ $SU(2)$ and Ref.~\cite{OP} for a general Lie group. In Ref.~\cite{DP1} a full
circle has been performed: starting from discretizing the $3d$ $SU(2)$ YM theory by the
lattice, making the duality transformation, going back to the continuum limit, one explicitly
recovers the partition function \ur{Z33}.

\section{Perturbation theory}

One may wonder how the usual perturbation theory with explicit (colour) gluons
at short distances looks like in the gauge-invariant formulation. Perturbation theory means
expansion in the coupling $g_{3,4}^2$. The curvature term is $O(1)$ in the coupling,
therefore in the leading order it has to vanish! In $3d$ it is just the ordinary scalar curvature
which has to vanish; in $4d$ it is a somewhat more complicated action $S_2$.
In $3d$, zero curvature means that the dual space is flat, which means that it is possible to
parametrize it by 3 out of the general 6 external coordinates $w^a$. [Three functions are just
a change of variables, like going from Cartesian to spherical coordinates -- it
does not make a flat space curvy.]  \\

In the gauge-invariant language, perturbation theory is expansion around the flat 
dual space.

\subsection{Where are gluons?}

Let us consider the simplest case of the $3d$ $SU(2)$ theory. If the dual space is flat, the 
\ae ther term is just the kinetic energy of three massless scalar fields $w^a$. This is exactly
what should be expected: the gauge-invariant content of the pertubation theory is one
transversely polarized gluon (times three colours).  In $3d$ there is only one transverse
polarization which can be described by a scalar field.  

In $4d$ case, the gauge-invariant content of the perturbation theory
are also transverse gluons, in this case there are two transverse polarizations,
times three colours. It can be shown that these degrees indeed show up
if one requires the curvature term $S_2$ to vanish \cite{DPBF}. 

\subsection{Where is the Coulomb force?} 

To explore the potential energy of two static quarks, one has to study the Wilson loop. 
The presence of the Wilson loop does not allow the parametrization 
of the metric tensor $g_{ij}$ by gradients $\partial_iw^\alpha$ as in \eq{gw} ; a {\it curl} 
must be added: 

\beq 
g_{ij}=(\partial_iw^\alpha+B_i)(\partial_jw^\alpha+B_j),\qquad 
{\rm Curl}\,{\bf B} = {\bf j}, 
\la{gwB}\eeq 
where ${\bf j}$ is the current along the Wilson loop, 

\beq 
{\bf j}=\int\!d\tau\,\frac{d{\bf x}}{d\tau}\,\delta({\bf x-x}(\tau)). 
\la{j}\eeq 
The leading order of the perturbation theory corresponds to the flat dual space, 
hence only three $w$'s are nonzero.  We thus get from the \ae ther term the Coulomb
interaction, plus transverse gluons \cite{DP4}: 
\bea 
\nonumber 
<W>&=&\int \!Dw^a\,\exp\left[ -\frac{g_3^2}{2}\int\! d^3x\, 
(\partial_iw^a+B_i)^2\right]\\ 
\nonumber 
&=&\det^{-\frac{3}{2}}(-\partial^2)\,\exp\, \frac{g_3^2}{2}\int\!d^3x\,B_i\, 
\frac{\delta_{ik}\partial^2-\partial_i\partial_k}{\partial^2}\,B_k \\ 
&=&\det^{-\frac{3}{2}}(-\partial^2)\,\exp \int\!d{\bf x}d{\bf y}\; 
{\bf j}({\bf x})\,\frac{g_3^2}{8\pi|{\bf x-y}|}\,{\bf j}({\bf y}), 
\la{Coul}\eea 
which is nothing but the Coulomb interaction between segments of the  
Wilson loop, as it should be in perturbation theory.

\section{A quick (but wrong) way to get confinement}

The aim of rewriting the YM theory in gauge-invariant variables
is to get new insights into the nonperturbative regime of the theory. 
One bonus can be obtained immediately by solving the classical
equations. We take the simplest case of the $3d$ $SU(2)$ theory \ur{Z32}
and consider the usual Wilson loop as a source. Varying the 
action of \eq{Z32} with respect to the metric we get the classical 
Einstein's equation modified by the presence of the \ae ther term
and the stress-energy tensor corresponding to the Wilson loop:

\beq
R^{ij}-\half\,R\,g^{ij}-\frac{ig_3^2}{\sqrt{g}}\,\delta^i_j=iT^{ij}.
\la{Einst}\eeq
The \ae ther term breaks the general covariance, of course. This equation 
can be solved for a large loop lying in a plane, and one gets for 
the average of the Wilson loop \cite{DP4}

\beq
<W>=\exp\left(-{\rm Area}\,g_3^2\,\delta(0)\right)
\la{W1}\eeq
{\it i.e.} the area law but with an infinitely thin string, because
it is a classical calculation. Quantum fluctuations are to smear out
the thin string and yield a finite string tension. Nevertheless, it is
amusing that in the dual formulation the first thing one gets is the 
area law!

\section{Order-of-magnitude analysis}

The reformulation of the YM theory in terms of gauge-invariant dual 
variables is quite unusual and to get some insight into the ensuing 
quantum theory, let us first of all analyze the order of magnitude 
of the fields and of their derivatives. We use the partition function
\ur{Z33}. Fluctuations leading to the action much greater than unity
are usually cut out. 

Let $a$ be some UV cutoff, {\it e.g.} the lattice spacing. From the
first (\ae ther) term we estimate: 

\beq
a^3g_3^2(\partial w)^2 \sim 1 \Rightarrow 
\partial w  \sim a^{-1}\,(g_3^2a)^{-\half},
\la{est1}\eeq
hence $\sqrt{g}\sim (\partial w)^3 \sim a^{-3}(g_3^2a)^{-\frac{3}{2}}$. 
The second (Einstein--Hilbert) term must be also of the order of unity (or less), 
otherwise the configuration of the field $w^\alpha(x)$ will be damped by oscillations from
nearby fluctuations. We have therefore:

\beq
a^3R\sqrt{g}\sim 1 \Rightarrow R \leq (g_3^2a)^{\frac{3}{2}} \to 0
\quad {\rm (!)}
\la{est2}\eeq
Unlike pure quantum gravity, the \ae ther term 
requires that quantum fluctuations of the metric should correspond to
zero curvature $R$ in the limit of vanishing lattice spacing! 
[A similar analysis applies to the $4d$ dual theory, with the only difference
that the curvature has to vanish not as a power but as one-over-logarithm
of the cutoff.] Does it mean that only small fluctuations around flat space are allowed,
leading to perturbation theory? Not necessarily. Non-zero values of $R$
are not prohibited provided  $R\sqrt{g}$ is a {\em full derivative.} If so,
large values of $R\sqrt{g}$ cancel between neighbour points, and the 
oscillating factor, $\exp\left(i\int R\sqrt{g}\right)$, does not damp such metrics. 
  
An example of a non-flat metric with $R\sqrt{g}$ being a full derivative is
the (anti) de Sitter $S^3$ space. It can be parametrized by 4 (out of the general~6)
external coordinates $w^A$ lying on a sphere, $\sum_{A=1}^4w^Aw^A=\frac{6}{R}$. 
$R\sqrt{g}$ becomes then the winding number of the map $S^3\mapsto S^3$,
which {\it is} a full derivative~\cite{DP2}. In the next section I show that dual space of
constant curvature seems to be, indeed, dynamically preferred.

\vspace{1cm} 
\section{Beyond perturbation theory: de Sitter dual space} 

Yet another parametrization of the curved dual space is obtained by expressing 
three external coordinates through the three other and introducing three
functions $U^\alpha=U^{1,2,3}(w^1,w^2,w^3)$ such that

\beq
g_{ij}=\partial_iw^a\partial_jw^bG_{ab}(w),\qquad
G_{ab}(w)=\delta_{ab}+\frac{\partial U^\alpha}{\partial w^a}
\frac{\partial U^\alpha}{\partial w^b}.
\la{gU}\eeq
[It is similar to describing the $2d$ surface by a function $z(x,y)$]. 
The \ae ther term becomes now the Lagrangian of a generalized $\sigma$
model:

\beq
{\cal L}=\partial_iw^a\partial_iw^bG_{ab}(w).
\la{sigma}\eeq
In $2+\epsilon$ dimensions integrating out frequences
from $\mu_1$ to $\mu_2$ renormalizes the metric as follows \cite{F}:

\beq
\frac{dG_{ab}}{d\ln\frac{\mu_2}{\mu_1}}=\epsilon\, G_{ab}
-\frac{1}{2\pi}R_{ab}-\frac{1}{8\pi^2}R_{ac}G^{cd}R_{db}-\ldots
\la{Fr}\eeq
where $R_{ab}$ is the Ricci tensor built from the metric $G_{ab}$ \ur{gU}.
This renormalization-group equation means that the metric gets renormalized as
one integrates out high frequences in quantum fluctuations. Putting boldly 
$\epsilon=1$ one discovers that the infrared-fixed point 
({\it i.e.} the zero of the r.h.s.) is the de Sitter space with curvature
$R=6\pi$ (to one loop accuracy) or $R=0.71\cdot 6\pi$ (in 2 loops), etc. 
Therefore, at low momenta the theory is described by the
$O(n)\;\sigma$-model with $n=4$, given by the partition function
\bea
\nonumber
Z &=& \int Dw^A\delta\left(w^2-\frac{6}{R}\right)\,
\exp\left(-\!\int\! d^3x\, \frac{g_3^2}{2}\partial_iw^A\partial_iw^A\right)\\
&=& \!\!\int \!D\lambda\! \int\! Dw^A\,\exp\int\!
d^3x\left[\!-\frac{g_3^2}{2} \left(\partial_i w^A\partial_i w^A + \lambda w^Aw^A\right)\!-\!
\frac{3\lambda g_3^2}{R}\right].
\la{Z34}\eea

\subsection{Spontaneous mass generation}

There is a well-known mechanism of spontaneous mass generation in the $O(n)$
$\sigma$-model through the Lagrange multiplier $\lambda$ getting a nonzero v.e.v. 
Strictly speaking, the mechanism is justified at large $n$ but it is plausible that
$n=4$ is large enough for this mechanism to work.

If it does, the four gauge-invariant fields $w^A$ obtain the mass 
$m_w^2 = <\!\lambda\!>\sim g_3^4$. 
This mass would be observable had we made lattice simulations in the dual formulation
\ur{Z33}. However, lattice simulations have been so far done in the usual formulation
implying that correlators of only quadratic, quartic,... combinations of the elementary
$w$ fields could be measured. Therefore, the observed glueballs are bound states
of the $w$ fields. From the quantum numbers, one can infer that the lightest glueballs'
masses are
\bea
\la{m2}
m_{0^{++}},\;m_{2^{++}}&=&2m_w\mp O\left(\frac{1}{n}\right),\\
\la{m4}
m_{0^{-+}},\;m_{1^{-+}}&=&4m_w\mp O\left(\frac{1}{n}\right), \qquad n=4,
\eea
since the $0,2^{++}$ operators can be constructed from two $w$ fields while
the $0,1^{-+}$ ones from minimally four. The correction is due to the interaction
suppressed as $1/n$. This seems to be in qualitative agreement with the arrangement 
of the lightest glueballs in $d\!=\!3$. Indeed, according to Teper \cite{T}, 
$m_{0^{++}}\approx 4.7\sqrt{\sigma}$,  $m_{2^{++}}\approx 7.8\sqrt{\sigma}$,  
$m_{0^{-+}}\approx 10\sqrt{\sigma}$,  $m_{0^{++}}\approx 11\sqrt{\sigma}$. 
Of course, one needs to compute the $1/n$ corrections before drawing
conclusions.    

\subsection{Area law}

The variables $w^A(x)$ have the meaning of dual gluons (but gauge-invariant!).
Mass generation for dual gluons usually implies confinement. Indeed, 
the average of a large Wilson loop is obtained by adding the mass term into \eq{Coul}: 
\bea 
\nonumber 
<\!W\!>&=&\int\!Dw^A\,\exp\left\{-\frac{g_3^2}{2}\int\! d^3x\,\left[
(\partial_iw^A+B_i)^2+m_w^2\,w^Aw^A\right]\right\} \\ 
&=&\exp\left\{ \frac{g_3^2}{2}\int\!d^3x\,
B_i\left(\!\delta_{ik}\,\partial^2-\frac{\partial_i\partial_k}{-\partial^2+m_w^2}\right)B_k\right\}. 
\la{Ar1}\eea
Recalling that ${\rm Curl}\,{\bf B} ={\bf j}$, (see \Eqs{gwB}{j}) and taking the magnetic field
created by a flat circle loop of radius $r$ to be $B_z=\theta(r-\rho)\delta(z),\; B_{x,y}=0$,
we obtain

\bea
\nonumber
 <\!\!W\!\!> &\stackrel{r\to\infty}{\longrightarrow} & \exp(-\sigma\,{\rm Area}),\\
\sigma &=& \frac{3}{32}\,g_3^2\,m_w.
\la{Ar2}\eea
Estimating $m_w$ from \Eq{m2} as half of the average of $0^{++}$ and $2^{++}$
masses we get the string tension $\sigma\approx 0.3\,g_3^2$ which is not so far from the
lattice value $0.335\,g_3^2$ \cite{T}. Given our negligence of  the expected $1/n=25\%$
corrections to all these formulae, the results for the glueball masses and the string
tension are not so bad. 

\section{Conclusions}

\noindent
1.  One can rewrite the quantum YM theory exactly in terms
of gauge-invariant variables. In $3d$ $SU(2)$ these are the
six coordinates $w^{1\!-\!6}(x)$ describing the embedding
of the curved dual space into flat space. The theory becomes Einstein's 
quantum gravity, plus the `\ae ther' term.  \\

\noindent 
2. Perturbation theory corresponds to expanding about flat dual space.
In the lowest order one recovers the Coulomb interaction, plus transverse
gluons.  \\ 

\noindent 
3. The probable infrared regime of the $SU(2)$ theory is that the dual space 
becomes a $S^3$ sphere described at low momenta by the $O(4)\,\sigma$-model.
In turn, it seems to generate spontaneously the mass for dual gluons, which gives rise to
glueball masses and the area law. \\

\noindent 
4. In $4d$ and/or for $SU(N)$ gauge groups the gauge-invariant variables are also 
related to the metric of the dual space. The larger $N$, the higher spin fields are needed  
to accommodate the necessary number of degrees of freedom of the original YM theory. 
The action consists of two terms: one has a new type of local symmetry mixing fields with
different spins, the other breaks this symmetry but is simple.  It is tempting to use this
formalism as a starting point for deriving a string from a local field theory. 

\vspace{.2cm}

\section*{Acknowledgments}
I am grateful to Victor Petrov for a collaboration and many useful discussions.
I would like to thank cordially Nora Brambilla and Giovanni Prosperi for inviting
me to Gargnano and for creating a wonderful atmosphere during the 
{\it Confinement-5} meeting.

\end{document}